\documentclass[12pt,a4paper]{article}
\usepackage[utf8]{inputenc}
\usepackage[english]{babel}
\usepackage{fullpage}
\usepackage{amsmath}
\usepackage{amssymb}
\usepackage{indentfirst}
\usepackage{epsfig}
\usepackage{cite}

\def\icarus{{Icarus}}           
\def\apj{{Astrophys. J.}}
\def\nat{{Nature}}
\def\apjl{{Astrophys. J. (Letters)}}
\def\planss{{Planet. Space Sci.}}
\def\apss{{Astrophys. and Space Sci.}}
\def\aap{{Astron. and Astrophys.}}
\def\mnras{{MNRAS}}

\title{Types of Gaseous Envelopes of ``Hot Jupiter'' Exoplanets}

\author{D. V. Bisikalo$^{1,2}$, P. V. Kaygorodov$^1$, D. E. Ionov$^1$, and V. I. Shematovich$^1$}
\date{}

\begin{document}

\newcommand{\Lp}[1]{${\mathrm L_{#1}}$}
\renewcommand{\vec}[1]{{\bf #1}}

\maketitle

\begin{center}
{\small\it $^1$Institute of Astronomy, Russian Academy of Sciences, Moscow, Russia}\\
{\small\it $^2$bisikalo@inasan.ru}
\end{center}

\begin{abstract}
As a rule, the orbital velocities of ``hot Jupiters,'' i.e., exoplanets with masses comparable to the mass of Jupiter and orbital semi-major axes less than 0.1 AU, are supersonic relative to the stellar wind, resulting in the formation of a bow shock. Gas-dynamical modeling shows that the gaseous envelopes around ``hot Jupiters'' can belong to two classes, depending on the position of the collision point. if the collision point is inside the Roche lobe of the planet, the envelopes have the almost spherical shapes of classical atmospheres, slightly distorted by the influence of the star and interactions with the stellar-wind gas; if the collision point is located outside the Roche lobe, outflows from the vicinity of the Lagrangian points \Lp1 and \Lp2 arise, and the envelope becomes substantially asymmetrical. The latter class of objects can also be divided into two types. If the dynamical pressure of the stellar-wind gas is high enough to stop the most powerful outflow from the vicinity of the inner Lagrangian point \Lp1, a closed quasi-spherical envelope with a complex shape forms in the system. If the wind is unable to stop the outflow from \Lp1, an open aspherical envelope forms. The possible existence of atmospheres of these three types is confirmed by 3D numerical modeling. Using the typical ``hot Jupiter'' HD~209458b as an example, it is shown that all three types of atmospheres could exist within the range of estimated parameters of this planet. Since different types of envelopes have different observational manifestations, determining the type of envelope in HD~209458b could apply additional constrains on the parameters of this exoplanet.
\end{abstract}

\section{Introduction}

``Hot Jupiters'' are currently the most intensively studied class of exoplanet. The reason is that these massive gaseous planets (with masses of the order of a Jupiter mass) are located in the immediate vicinity of the star (at distances not exceeding 0.1~AU~\cite{Murray-Clay-2009}). This makes it possible to obtain the most reliable observational data. Precisely these objects were the first exoplanets discovered~\cite{Mayor-Queloz-1995}, and therefore they have been studied the longest. In principle, the most promising objects for detailed studies are transiting exoplanets, i.e., planets whose inclination (the angle between the line of sight and orbital plane) is so small that it is possible to observe the planet when it crosses the stellar disk. In this case, combining radial-velocity measurements for the star and spectral and photometric observations of a transit, it is possible to determine both the mass and radius of the planet. Moreover, it is possible to obtain absorption spectra of the upper layers of the planetary atmosphere, which provide information on its structure and composition. The first transiting ``hot Jupiter'' HD~209458b was discovered in 1999~\cite{Charbonneau-2000}. Currently, about 300 transiting exoplanets are known and more than 50\% are ``hot Jupiters''.

Note that observations of ``hot Jupiters'' by the Hubble Space Telescope (HST) provide evidence for complex physical processes in the gaseous envelopes (atmospheres) of these planets. For example, the Ly-$\alpha$ absorption during the transit HD~209458b reaches 9\%-15\%~\cite{Vidal-Madjar-2003, Vidal-Madjar-2008, Ben-Jaffel-2007}, while the disk of the planet reduces the brightness of the star by only~1.8\%. This means that the planet is enshrouded by an extended neutral-hydrogen envelope whose size exceeds the Roche lobe of the planet. This was later confirmed by transit observations in lines of carbon, oxygen, and silicon~\cite{Vidal-Madjar-2004, Ben-Jaffel-2009, Linsky-2010}. The depth of the transit in lines of these species reached $8\%\div 9\%$. Two main hypotheses have been proposed to explain this. The first envisages expansion of the envelope due to heating by stellar radiation, resulting in outflow at a rate of $\sim 10^{10}$~g/s. Several one-dimensional atmosphere models with strong outflows of matter were computed to test this hypothesis~\cite{Yelle-2004, Munoz-2007, Koskinen-2012}. The second hypothesis is based on the possibility of charge exchange by ions of the stellar wind and atmospheric atoms~\cite{Lammer}. The atoms formed have high enough energy to overcome the gravitational potential of the planet, and, hence, can form an envelope around the planet.

The existence of large gaseous envelopes around ``hot Jupiters'' was confirmed by observations of HD~189733b~\cite{Lecavelier-2010} and WASP-12b~\cite{Fossati-et-al:2010, Fossati-et-al:2010b}. Moreover, observations of these planets led to the discovery of an even more interesting phenomenon -- an offset between the ingress or egress in different spectral ranges. This was first discovered by Fossati~et~al.~\cite{Fossati-et-al:2010} in 2009, who observed the transit of WASP-12b in the UV using the HST. UV observations were carried out in several 40~\AA-wide bands. According to these observations, in some light curves, the ingress starts in the UV about 50~min earlier than in the visual. This provides evidence for intervening absorbing material at a distance of about 4-5 planetary radii.

It was suggested that this matter may flow from the planet to the star~\cite{Lai-et-al:2010, Li-et-al:2010}, or that a bow shock has formed due to the intrinsic magnetic field of the planet~\cite{Vidotto-2010, Vidotto-2011, Vidotto3}. However, in the first hypothesis, the envelope is completely open and, for currently accepted parameters of WASP-12b, its lifetime should not exceed several years. The second hypothesis requires an appreciable magnetic field, which is hard to explain if, as is typical for ``hot Jupiters'', the planet rotates synchronously with the star. Bisikalo et al.~\cite{Bisikalo-2013} suggested that the early ingress could be due to the formation of a closed, aspherical envelope around WASP-12b that substantially exceeds the Roche lobe. This envelope forms due to the outflow of the planetary atmosphere through the vicinity of \Lp1 and \Lp2. The supersonic motion of the planet in the stellar-wind gas results in the formation of a complex bow shock, which constrains the outflow of the matter through the Lagrangian points and makes the envelope stationary and long-lived. Additional properties of the gaseous envelopes formed were derived from Ly-$\alpha$ observations of the late phases of the transit of the planet HD~189733b~\cite{Lecavelier-2012}. In particular, it was shown that the late egress is due to absorption in the flow directed toward the observer.

Our numerical modeling in the present study showed that the formation of aspherical envelopes is quite common for the ``hot Jupiters''. Moreover, it became clear that these envelopes can be both closed and open, i.e., conserving or losing mass, respectively. We carried out a series of computations of gas-dynamical interactions of the atmosphere of the exoplanet HD~209458b with the stellar wind for various atmospheric parameters found in the litera- ture, with the aim of finding criteria that differentiate the three types of envelopes -- closed, quasi-closed, and completely opened~\cite{Murray-Clay-2009, Munoz-2007, Koskinen-2010, Koskinen-2012}.

The paper is organized as follows. In Section~\ref{phys}, we consider analytically criteria defining the three types of envelopes, and suggest a method that can be used to estimate the type of envelope forming using the known parameters of an exoplanet. Section~\ref{results} describes our numerical model and presents the results of our 3D gas-dynamical computations. Section~\ref{concl} summarizes the main results, which confirm existence of three types of gaseous envelopes around ``hot Jupiters''.

\section{Physical basis for the classification of the atmospheres of ``hot jupiters''}
\label{phys}

Let us consider a star-planet system with a circular orbit and synchronously rotating components. We will also assume that the magnetic field is weak and does not appreciably influence the gas dynamics of the outer layers of the planetary atmosphere. In the absence of external forces, the atmosphere of the gas giant should be spherical. The atmospheres of ``hot Jupiters'', which are typically located close to their star, are influenced by several additional forces which give rise to deviations from a spherical shape. First and foremost, these are the forces described by the Roche potential -- the gravitational attraction of the components and the centrifugal force:
\begin{equation}
\Phi=-\frac{GM_*}{\sqrt{x^2+y^2+z^2}}-\frac{GM_{pl}}{\sqrt{(x-A)^2+y^2+z^2}}-
\frac{1}{2}\vec{\Omega}^2\left(\left(x-A\frac{M_{pl}}{M_{pl}+M_*}\right)^2+y^2\right),
\label{eqroche}
\end{equation}
where~$G$ is the gravitational constant, $M_{pl}$ the mass of the planet, $M_*$ the mass of the star, $\vec{\Omega}$ the angular velocity of the star-planet system, and $A$ the radius of the planet’s orbit. The Roche potential has five libration points at which its gradient is zero. The two located in the immediate vicinity of the planet are most important for studies of the atmosphere: \Lp1 and \Lp2. \Lp1, also called the inner Lagrangian point, is located between the planet and the star. The equipotential surface passing through \Lp1 delimits the Roche lobes of the planet and star. Since all forces in Eq.~\eqref{eqroche} are balanced at \Lp1, matter can flow freely from the planet to the star through \Lp1 under the action of the pressure gradient.

Another important effect in determining the shape and size of the atmosphere of a ``hot Jupiter'' is the interaction of the atmosphere with the stellar wind. Since the typical scales of the orbits of ``hot Jupiters'' are small, their orbital velocity is usually higher than the sound speed in the interplanetary medium, so that a bow shock should form. In this case, a gas-dynamical structure should form in the vicinity of the planet, consisting of this shock and a contact discontinuity separating the gas of the atmosphere and the stellar wind. The impact of the stellar wind can be assessed by considering momentum conservation, written for a tangential discontinuity~\cite{Landau}:
\begin{equation}
\rho_1 v_1^2+p_1=\rho_2 v_2^2+p_2,\label{press1}
\end{equation}
where~$\rho_1, \rho_2$ are densities, $v_1, v_2$ are velocities, and $p_1, p_2$ are the pressures on either side of the discontinuity. For a fixed, spherical, isothermal atmosphere located in the gravitational field of a point mass, the density and pressure depend on the radius $r$ as
\begin{equation}
\begin{array}{rcl}
\rho_{atm}(r)&=&\rho_0\cdot\exp\left\{-\dfrac{G\,\,M_{pl}}{R_{gas} T_{atm}}\left(\dfrac{1}{r_0}-\dfrac{1}{r}\right)\right\}\\[5mm]
p_{atm}(r)&=&\rho_{atm}\, R_{gas}\, T_{atm}
\end{array},\label{atmos}
\end{equation}
where~$\rho_{atm}$ is the density of the atmosphere at the radius $r$, $\rho_0$ the corresponding density at some radius $r_0$ (usually taken to be the photometric radius of the planet), $R_{gas}$ the gas constant, $T_{atm}$ the temperature of the planet, and $p_{atm}$ the atmospheric pressure at the radius $r$.

Substituting the density and atmospheric pressure into the left-hand side of~\eqref{press1}, and the density, pressure and velocity of the stellar wind incident onto the atmosphere into the right-hand side of~\eqref{press1}, we obtain an equation similar to the equation used to determine the shape of the atmosphere in~\cite{Baranov}:
\begin{equation}
p_{atm}(r)=\rho_w \vec{v_w}^2\cos^2(\vec{n},\vec{v_w})+p_w,\label{barkras}
\end{equation}
where~$\rho_w$ is the wind density, $\vec{v_w}$ the velocity of the wind, and $\vec{n}$ the vector normal to the surface of the atmosphere. Equation~\eqref{barkras} can be used to find the shape of the windward side of the atmosphere, directly interacting with the stellar wind. The head-on collision point (HCP), where $\cos(\vec{n},\vec{v_w})=1$, is located closest to the center of the planet.

\begin{figure}
\begin{center}
\epsfig{file=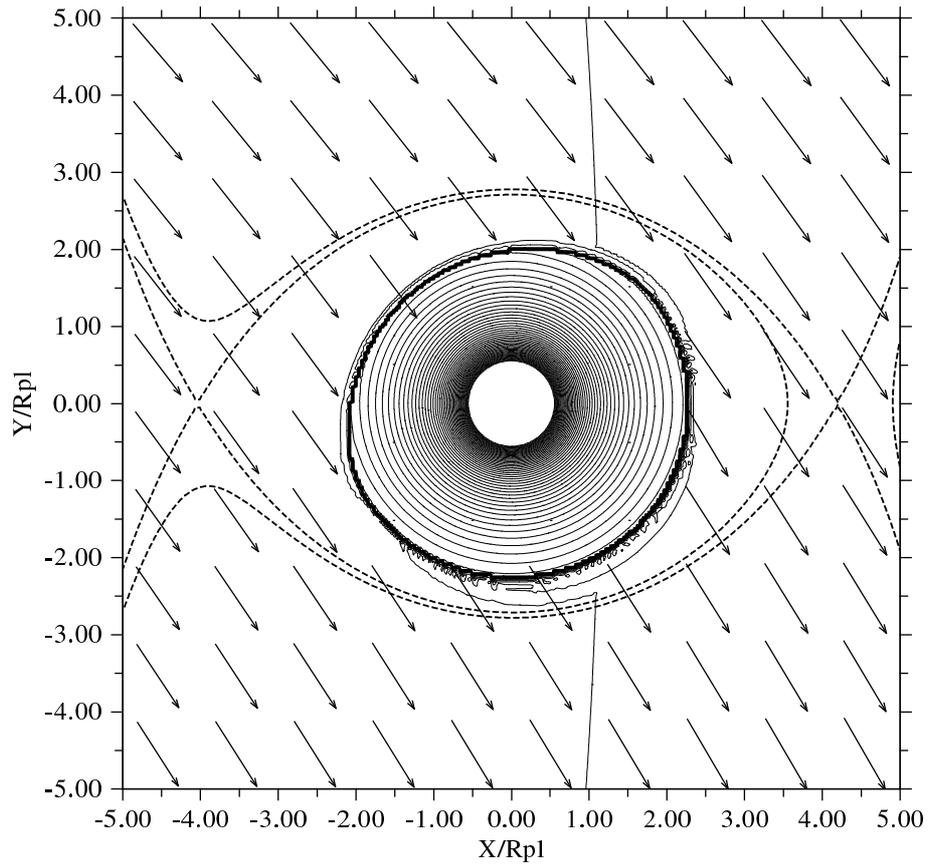, width=12cm}
\end{center}
\caption{The shape acquired by the atmosphere of a ``hot Jupiter'' under the action of the stellar wind. The velocity field and contour lines of the density in the equatorial plane are shown. The dashed lines show contours of the the Roche potential.}\label{zerot}
\end{figure}

The shape of the atmosphere, distorted by the wind pressure, is shown in Fig.~\ref{zerot}. The stellar-wind velocity field in Fig.~\ref{zerot} is shown in the reference frame rotating with the star-planet system. Since the velocity of the orbital motion of the planet is added to the radial velocity of the wind, the flow incident onto the atmosphere will be inclined relative to the axis connecting the system components. Accordingly, the symmetry axis of the atmosphere will also be tilted.

\begin{figure}[p]
\begin{center}
\epsfig{file=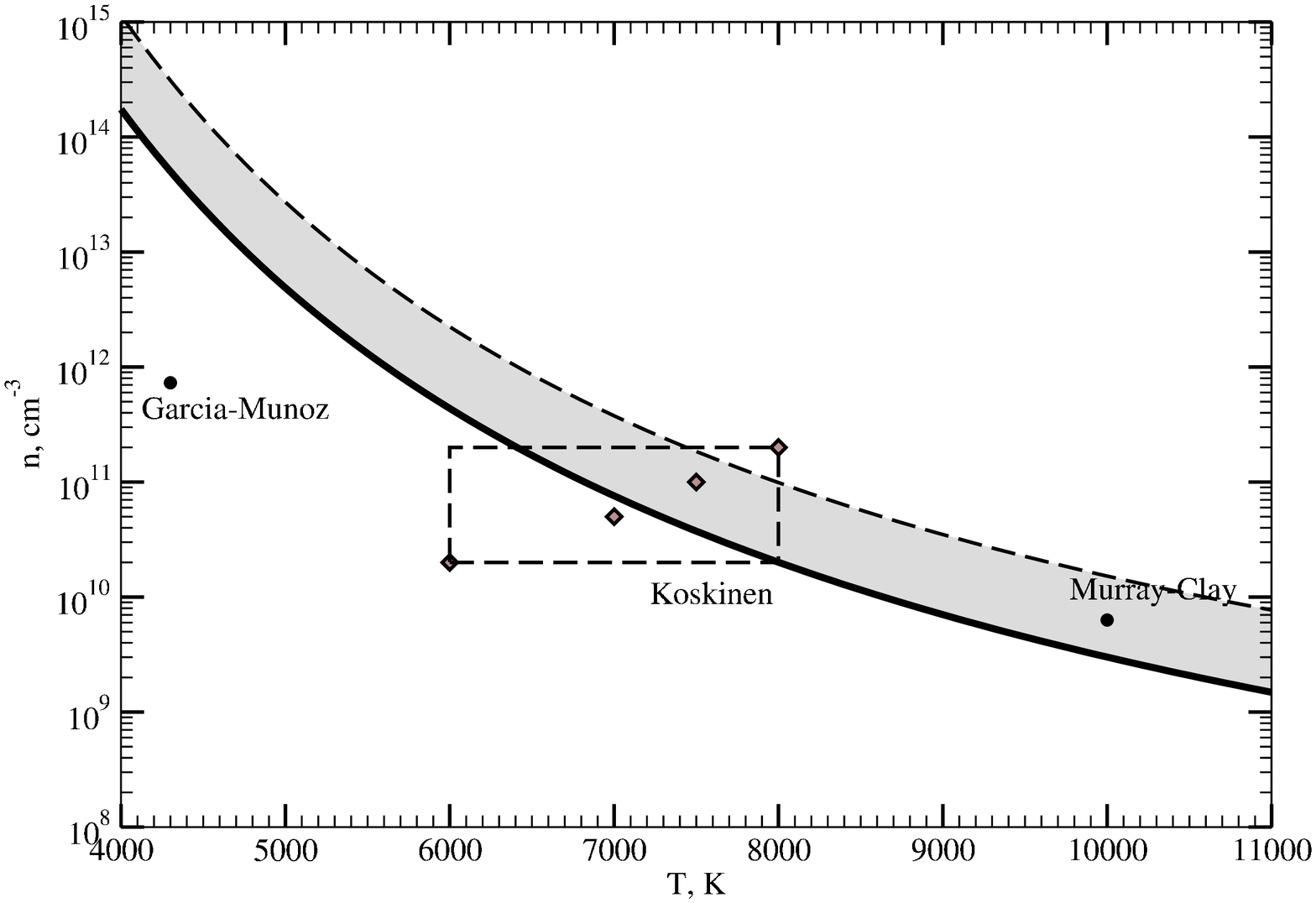, width=14cm}
\end{center}
\caption{Parameters of the atmosphere of HD~209458b (temperature and number density of the matter at the photometric radius) for which the planet’s atmosphere can correspond to the one of the three identified types. Below the solid line are the points for which the atmosphere will be closed lie below the solid curve. The existence of a quasi-closed envelope in which the flow from \Lp1 is arrested by the stellar wind is possible in the shaded area. The region of open atmospheres lies above the dashed curve. The dots show the values obtained in~\cite{Munoz-2007} and~\cite{Murray-Clay-2009}. The rectangular area corresponds to the range of values given in~\cite{Koskinen-2010} for the upper atmosphere. The diamonds show the parameters for which 3D modeling was carried out in the present paper.}\label{rhot}
\end{figure}

As the pressure of the atmosphere depends on the density $\rho_0$ and the temperature, varying these parameters, it is possible to obtain either an atmosphere that is located completely inside the Roche lobe of the planet, or that partially extends beyond the Roche lobe. Solving~\eqref{barkras} for various values of $\rho_0$ and $T$, it is possible to define parameters for which the HCP of the atmosphere and stellar wind will be inside or outside the Roche lobe. When the HCP is located inside the Roche lobe, the atmosphere is closed. When this is not the case, outflows can form. The solid curve in Fig.~\ref{rhot} separates the regions of completely closed (below the curve) and outflowing (above the curve) atmospheres calculated for the explanet HD~209458b using stellar-wind parameters similar to those for the solar wind~\cite{Withbroe-et-al:1988}.

If the atmospheric parameters correspond to the region above the solid curve in Fig.~\ref{rhot}, outflows from the vicinity of the inner Lagrangian point \Lp1 can form\footnote{Since the points \Lp2 and \Lp1 arise almost synchronously with changes in the atmospheric parameters, we will discuss the criteria for the opening or closing of \Lp1 only.}. As was shown in~\cite{Bisikalo-2013}, such outflows (flows from the points~\Lp1 and~\Lp2) can sometimes be stopped by the dynamical pressure of the stellar wind. We can theoretically estimate the distance where the flow from \Lp1 can be stopped and a criterion for this to occur using~\eqref{press1}, substituting the flow parameters (density~$\rho_s$, velocity~$\vec{v_s}$, and pressure\footnote{The assumption that the flow has the same temperature as the atmosphere along its entire extent is acceptable, because the heating of the flow gas should proceed similarly to heating of the outer parts of the planetary atmosphere.} $p_s=\rho_s\cdot R_{gas}\cdot T_{atm}$) into one side and the parameters of the wind into the other. Since we are interested in a criterion for the flow to be stopped (not deflected), we will consider only those points along the flow where a HCP can exist, i.e., the the velocities of the flow and the wind are collinear.

\begin{figure}
\begin{center}
\epsfig{file=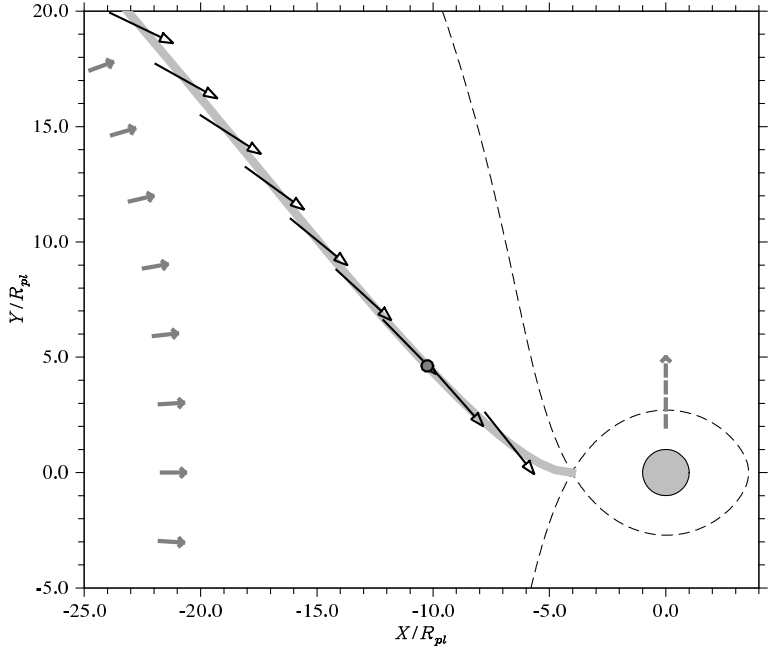, width=10cm}
\end{center}
\caption{Ballistic trajectory of the flow from \Lp1 (bold gray line). The dashed line shows the contour of the Roche potential passing through \Lp1. The center of the planet is located at (0, 0). The arrows crossing the path of the streamline indicate the direction of the stellar wind in a reference frame rotating with the star-planet system. The solid gray arrows show the direction of the radial motion of the wind. The dashed gray arrow shows the direction of the orbital motion of the planet. The dark gray circle on the ballistic trajectory corresponds to the point where the flow direction is collinear to the direction of the stellar wind.}\label{scheme}
\end{figure}

Figure 3 shows schematically the ballistic trajectory of the flow from \Lp1. The arrows crossing the path of the flow indicate the directions of gas streams in the stellar wind at the corresponding points of the flow. At a certain point (indicated by the light gray circle), the direction of the flow is collinear to the stellar-wind velocity. This makes it possible to solve~\eqref{press1}. For the exoplanet HD~209458b, this point is at a distance of~$\sim 7 R_{pl}$ from~\Lp1. This position corresponds to the position of the bow shock that gives rise to the early ingress in the UV, at a distance of~$\sim 4.8 R_{pl}$ ahead of the planet projected onto the stellar limb, remarkably consistent with observations of the early ingress in the WASP-12b system~\cite{Fossati-et-al:2010}. For the adopted wind parameters, the density of the flow at this point must be~$\sim 10^{-18} \text{g}/\text{cm}^3$ to stop the flow. If the density of the flow exceeds this value, the flow will continue to move outward and cannot be stopped, since there will be no points where the flow velocity and wind velocity will be collinear along its subsequent path. If the flow density is below the critical value, the flow can be deflected, and the HCP will move to the position where the condition~\eqref{press1} is satisfied.

Let us determine the criteria under which a flow of matter cannot be stopped by the dynamical pressure i.e., the planet efficiently (over several years) loses its atmosphere. To do this, we must compute the critical density~$\rho_0^*(T_{atm})$, distinguishing solutions with quasi-closed atmospheres ($\rho_0<\rho_0^*$) corresponding to outflows that are stopped by the stellar wind, leading to the formation of extended aspherical envelopes, from solutions with open atmospheres, for which~$\rho_0>\rho_0^*$. These calculations must take into account several important physical effects. Since the flow is accelerated by the gravitational field of the star, the flow density should decrease toward the star. We can use the known equations for flow acceleration~\cite{Lubow-Shu:75} and the Bernoulli equation to determine the rate of decrease of the flow density during the flow propagation. Assum- ing that the flow density at \Lp1 is the equilibrium density of the atmosphere, we can calculate the density at the photometric radius~$\rho_0$, knowing the density at the point where the flow stops. Since the value of the critical density depends on the adopted temperature of the atmosphere [see~\eqref{atmos}], the final critical density should be searched for in the form~$\rho_0^*(T_{atm})$.

The critical density corresponding to quasi-closure of the atmosphere for the exoplanet HD~209458b is shown by the dashed line in Fig.~\ref{rhot}. All points located in the shaded area between the two curves correspond to the solutions with quasi-closed atmospheres. Essentially all the atmospheric parameters obtained for HD~209458b are located in the area of closed or quasi-closed atmospheres, within the parameter uncertainties~\cite{Koskinen-2010}.

\section{Simulation results for atmospheres of various types}
\label{results}

\begin{table}
\begin {center}
\begin {tabular} {rcl}
\hline
\hline\\
Mass of the planet  & 0.64 & $M_{Jup}$\\
Mass of the star  & 1.1 & $M_{\odot}$\\
Radius of the planet & 1.32 & $R_{Jup}$ \\
Radius of the star & 1.1 & $R_\odot$ \\
Semi-major axis of the orbit & 0.045 & A.U.\\
Orbital period & 3.5 & day \\[5mm]
\hline
\hline
\end{tabular}
\end{center}
\caption{Parameters of the~HD 209458 system~\cite{Linsky-2010}}
\label{tab1}
\end{table}

The computations were carried out using the numerical model described in~\cite{Bisikalo-2013}. The parameters of the components of the HD~209458 system are listed in Table~\ref{tab1}. The velocity of the motion of the planet in the system is 141~km/s. For the adopted stellar-wind parameters, similar to the solar-wind parameters at the same distance from the star, $T_w=7.3 \cdot 10^5$ K, $n_w=1.4 \cdot 10^4\, \text{cm}^{-3}$, $v_w=100$~km/s~\cite{Withbroe-et-al:1988}, the orbital motion of the gas is supersonic, with a Mach number of~$M=1.4$.

As in~\cite{Bisikalo-2013}, the flow is described by a 3D~system of gasdynamical equations closed with the equation of state for a neutral, ideal monatomic gas. Nonadiabatic radiative cooling of the gas was not included. The system of gas-dynamical equations was solved using a Roe-Osher TVD scheme with the Einfeldt modification. This method is described in detail in~\cite{Boyarchuk-et-al:2002, Bisikalo-et-al:2003}.

The calculations were performed in a rotating reference frame with the coordinate origin at the center of the star. To approximately include the acceleration of the stellar wind, the force field was described by a modified Roche potential:
\begin{equation}
\Phi=-\Gamma \frac{GM_*}{\sqrt{x^2+y^2+z^2}}-\Gamma \frac{GM_{pl}}{\sqrt{(x-A)^2+y^2+z^2}}-
\frac{1}{2}\Omega^2\left(\left(x-A\frac{M_{pl}}{M_*+M_{pl}}\right)^2+y^2\right) \,,
\end{equation}
where~$\Gamma$ -- is a parameter that is equal to 0 in regions filled by the stellar wind, and to 1 otherwise. This approach enabled us to keep the radial velocity of the stellar wind constant on the scales encompassed by the computational domain.

The calculations were made on a Cartesian grid. The coordinate origin was placed at the center of the star, the~$X$ axis was directed toward the planet, the~$Z$ axis coincided with the axis of rotation and was perpendicular to the orbital plane of the system, and the~$Y$ axis completed the right-handed system. The computational domain was $(20\times 20\times 10)$ $R_{pl}$ in size, and the grid resolution was $464\times 464\times 182$ cells. The grid became denser along the $X$ and $Y$ directions toward the center of the planet, with $q=1.011$. This coefficient was chosen to ensure that the density gradients at the inner boundary of the computational domain were not too high.

\begin{table}
\begin {center}
\begin {tabular} {c|cc}
Model&T, K & n, $\text{cm}^-3$\\[1mm]
\hline
\hline\\[-3mm]
1 & 6000 & $2\cdot10^{10}$\\
2 & 7000 & $5\cdot10^{10}$\\
3 & 7500 & $1\cdot10^{11}$\\
4 & 8000 & $2\cdot10^{11}$
\end{tabular}
\end{center}
\caption{Parameters of the atmosphere used in the modeling: temperature and number density at the photometric radius.}\label{tab2}
\end{table}

The parameters of the atmosphere were specified according to the latest estimates~\cite{Koskinen-2012}. The spread of these parameters is shown by the rectangle in Fig.~\ref{rhot}. For the calculations we selected four pairs of parameters covering the entire area along a diagonal. The model parameters are listed in Table~\ref{tab2}. This set of parameters enables the construction of all three types of atmospheres suggested by analytical considerations.

\begin{figure}
\begin{center}
\epsfig{file=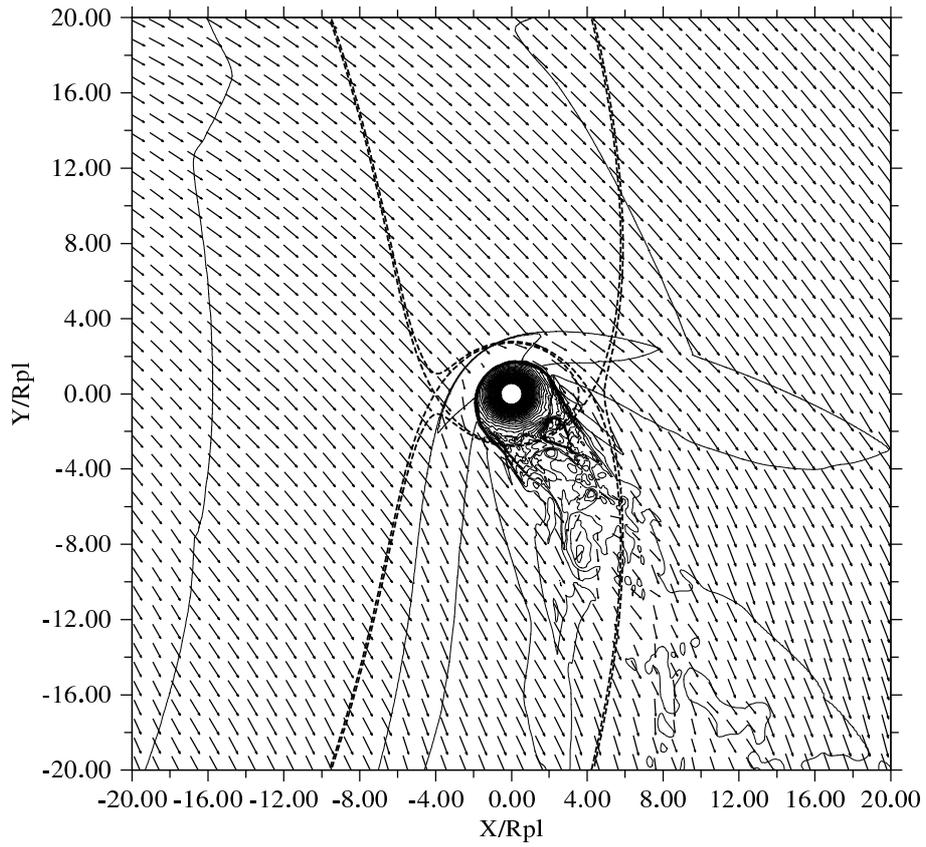, width=12cm}
\end{center}
\caption{Results of computations for Model~1. Contour lines of the density and velocity vectors in the equatorial plane of the system are shown. The center of mass of the planet is at (0, 0); the units of the distance are~$R_{pl}$. The dashed curves show the equipotential lines of the Roche potential.}\label{fig2}
\end{figure}

\begin{figure}
\begin{center}
\epsfig{file=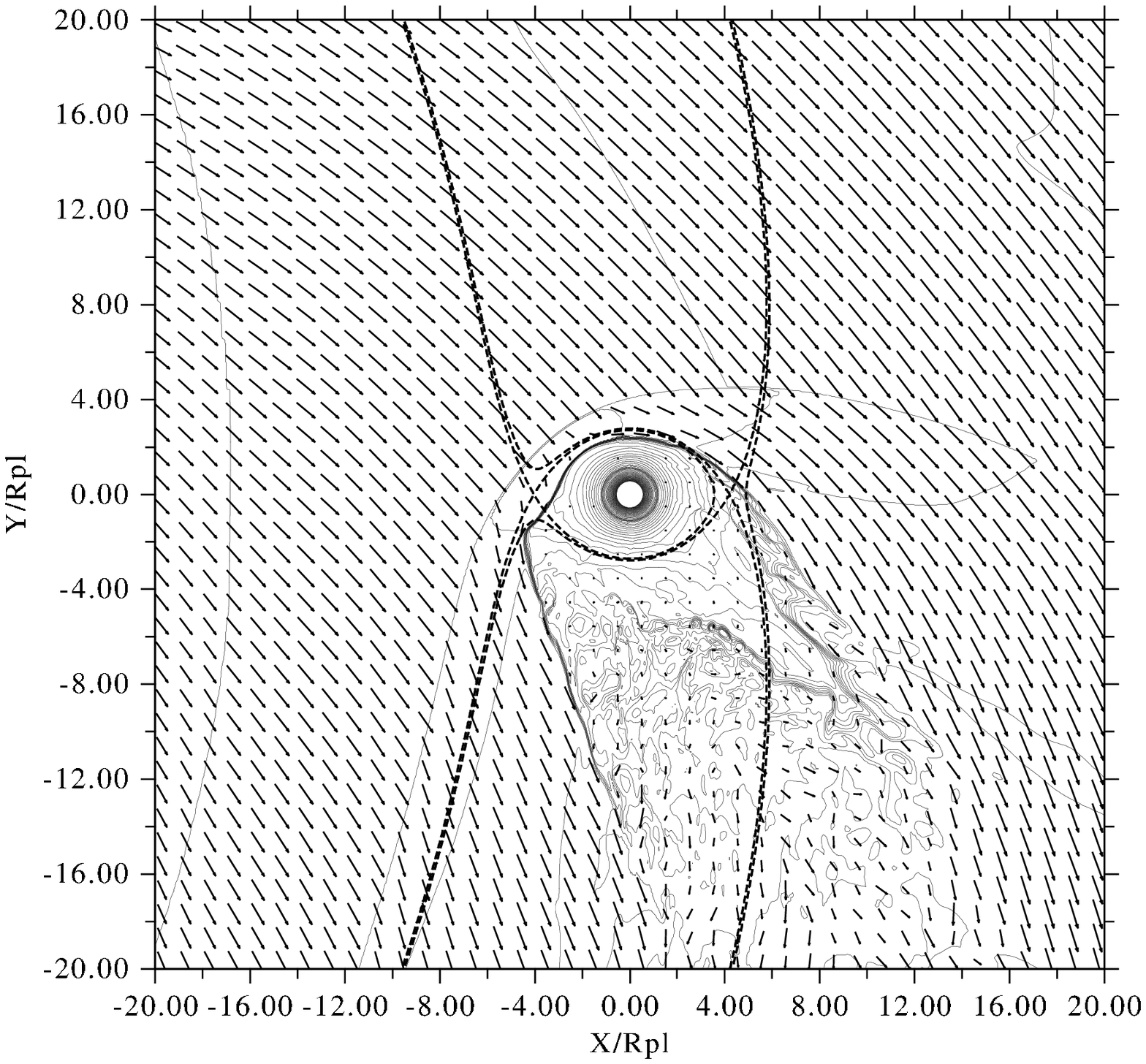, width=12cm}
\end{center}
\caption{Same as Fig.~\ref{fig2} for Model 2.}\label{fig3}
\end{figure}

\begin{figure}
\begin{center}
\epsfig{file=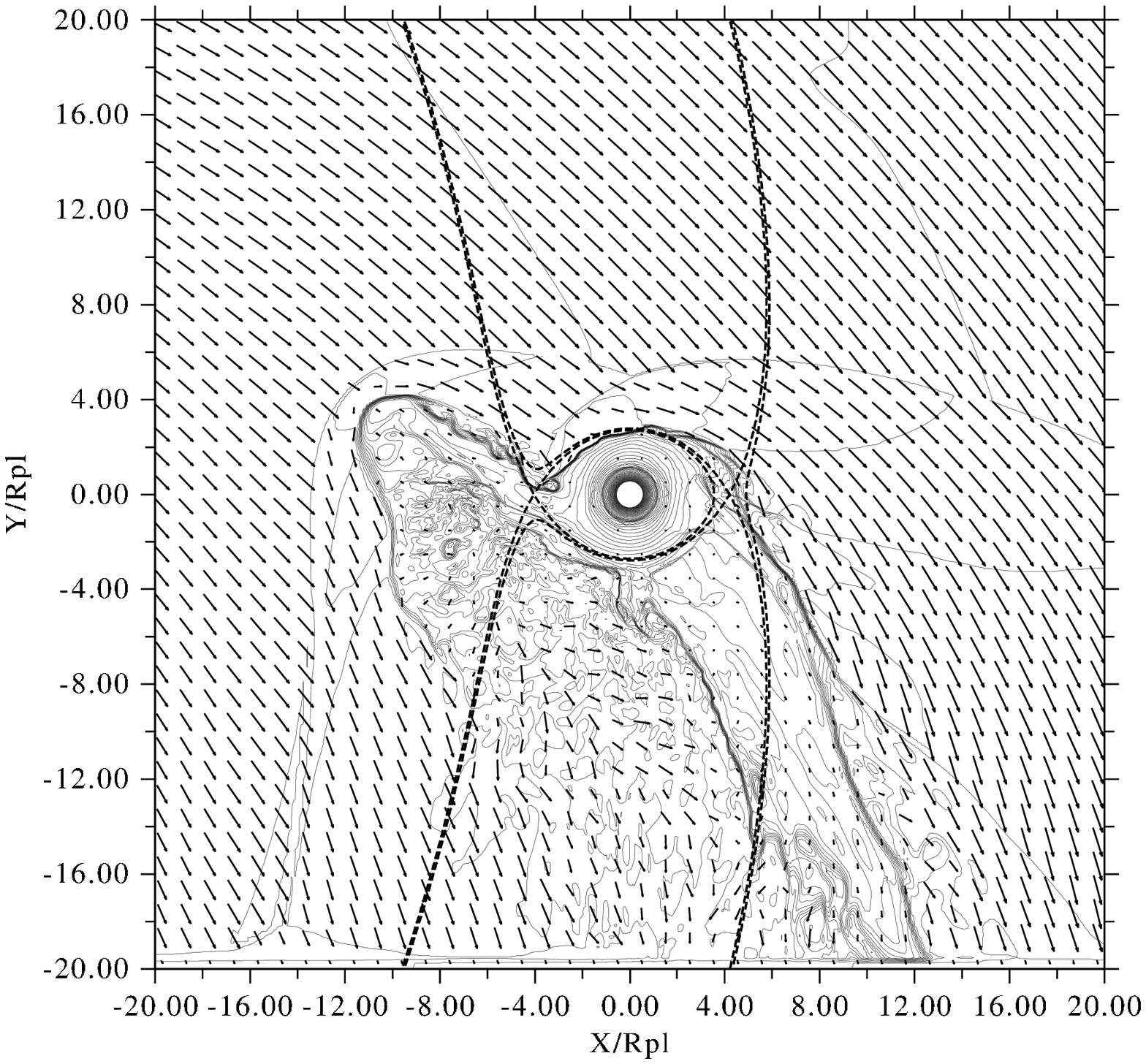, width=12cm}
\end{center}
\caption{Same as Fig.~\ref{fig2} for Model 3.}\label{fig4}
\end{figure}

\begin{figure}
\begin{center}
\epsfig{file=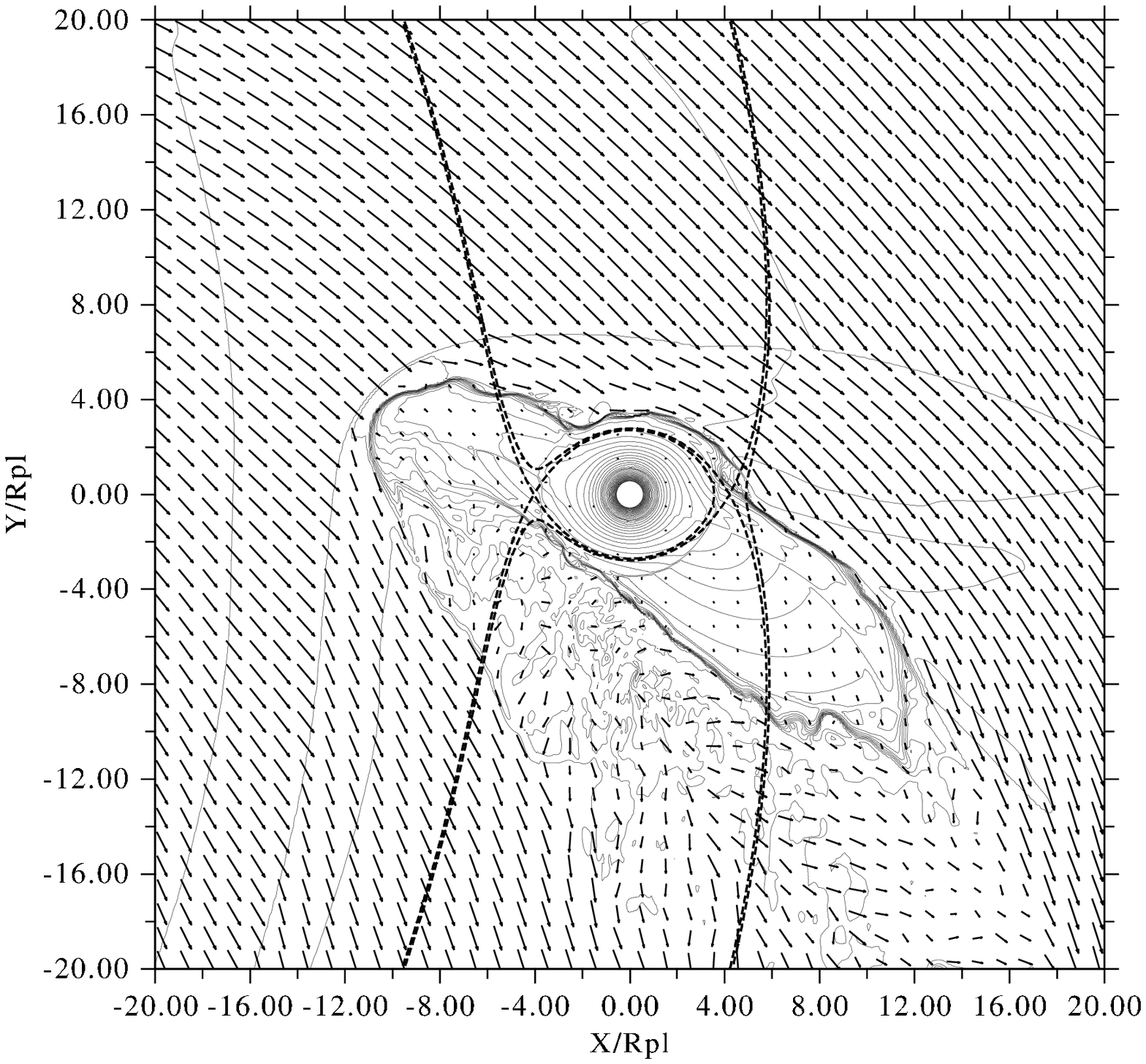, width=12cm}
\end{center}
\caption{Same as Fig.~\ref{fig2} for Model 4.}\label{fig5}
\end{figure}

Results of numerical simulations for four sets of adopted parameters are shown in Figs.~\ref{fig2}--\ref{fig5}. The resulting gas-dynamical patterns are very different.

In Model 1 (Fig.~\ref{fig2}), a closed atmosphere is obtained, with the stellar wind streaming around it. A symmetrical bow shock forms, with an approximately spherical shape near the HCP and tending to a Mach cone at some distance from this point. The contact discontinuity delimiting the size of the atmosphere is located completely inside the Roche lobe of the planet. In general, the shape of the planet’s atmosphere deviates only slightly from a sphere.

In Model 2 (Fig.~\ref{fig3}) the shape of the atmosphere is substantially aspherical. The HCP is further from the planet than in Model 1, but remains within the Roche lobe. Two bulges directed toward the points \Lp1 and \Lp2 are clearly visible. These lead to a pronounced change in the shape of the bow shock and contact discontinuity. The width of the tail is much larger than in Model 1. Interestingly, the matter does not flow toward the star from \Lp1, but there is a little outflow from the atmosphere through the vicinity of \Lp2. Thus, the atmosphere is partially open, although the analytical estimates suggested the atmosphere should be closed for this set of parameters (Fig.~\ref{rhot}). This indicates the estimated accuracy of the analytical estimates to be a few percent; this is understandable, since gas- dynamical effects were not taken into account.

Figures~\ref{fig4} and~\ref{fig5} show the results for Models 3 and 4. It is evident that the nature of the matter flows has changed qualitatively. Two powerful streams from the vicinity of \Lp1 and \Lp2 have formed. In contrast to the solutions typical for outflows in close binary stars~\cite{Bisikalo-et-al:2003}, the areas of formation of the flows have considerable dimensions. The flow from \Lp1 starts in a fairly large region between the Lagrangian point and the upper edge of the Roche lobe. The base of the flow from \Lp2 has roughly the same size. However, further along the flows, they become markedly different. While the flow from \Lp1 gradually narrows, the flow from \Lp2, on the contrary, expands significantly. The density contours clearly show that, at a given distance from the planet, the densityis much higher in the flow from \Lp1 than in the flow from \Lp2. Note that the flow from \Lp2 contains an additional shock along the stream, which probably originated as a result of the decay of the discontinuity at the boundary of the flow and the atmosphere.

The bow shock in this solution is a complex of two shocks, one formed around the atmosphere and the other around the stream from \Lp1. Another shock forms in the place where they intersect, extending from the intersection of the stream and atmosphere to a kink in the shock at the junction of its two branches. The HCP is located at the edge of the outflow from \Lp1. The bow shock in front of the atmosphere occupies a much larger fraction of the computational domain compared to the previous models. At the HCP, it extends to the boundary of the Roche lobe. A wide tail forms behind the planet. Since it forms both behind the planet and behind both flows, its width is much larger than the width of the tail in the previous models.

Despite the similar flow patterns in Models~3 and~4, they are fundamentally different. The solution in Model~3 is quasi-closed, i.e, the outflows from the Lagrangian points are partially stopped by the action of the stellar wind, and a closed envelope with a small outflow of matter along the discontinuity forms. The outflow from \Lp1 is not stopped in Model~4, and continues to propagate in the direction of the star, with all matter stored in it leaving the envelope, leading to a high mass loss. The fate of this matter is of interest, since the formation of a torus of dense matter or even of a disk is possible in such a system. Despite the numerical complexity of this problem, we plan to carry out the necessary modeling in later studies.

\section{discussion and conclusions}
\label{concl}

The analysis presented in this paper supports the idea that the gaseous envelopes around ``hot Jupiters'' can be classified accordind to three types. Most importantly, it is necessary to identify closed envelopes, formed in systems where the head-on collision of the flow with the stellar wind occurs inside the Roche lobe. Depending on the degree of filling of the Roche lobe, its shape may deviate from spherical, but they clearly lack significant outflows of matter. If the head-on collision point, and, therefore, some of the atmosphere, lie outside the Roche lobe, a significant outflow from the vicinity of the Lagrangian points \Lp1 and \Lp2 begins. In this case, some of the envelopes may be limited in size, since the flows arising due to Roche-lobe overflow can be stopped by the stellar wind. As a consequence, quasi-closed, stationary, long-lived, and appreciably aspherical envelopes form. If the wind can not stop the flows, an open, aspherical envelope forms in the system.

Clearly, this classification depends on the degree of Roche-lobe filling by the planetary atmosphere, i.e., on the temperature and density of the upper atmosphere. Since the orbits of ``hot Jupiters'' are close to their parent stars, their upper atmospheres are exposed to intense flows of plasma and radiation from the stars. As a result, these planets have dense and extended thermospheres and ionospheres. Preferential absorption of stellar radiation in the soft X-ray and extreme ultraviolet occurs in the upper atmosphere, considerably increasing the atmospheric temperature and possibly leading to the overflow of the planet’s Roche lobe. Aerodynamical models of the upper atmosphere of HD~209458b~\cite{Yelle-2004, Munoz-2007, Koskinen-2012} assume that, when temperatures of the order of several thousand Kelvin and densities of~$10^{10}-10^{11} \text{cm}^{-3}$ are reached, the thermal destruction of molecular hydrogen occurs, and a transitional $H_{2} \rightarrow H$ region forms in the lower thermosphere of the exoplanet. The photoionization of atomic hydrogen begins to play the dominant role in the upper thermosphere. Accordingly, the composition of the upper atmosphere varies with altitude as $H_{2} \rightarrow H \rightarrow H^+$, and this becomes an additional factor in the formation of an extended atmosphere, since this change is accompanied by an increase in the characteristic scale height, and the gas temperature reaches several thousand Kelvin.

Indeed, a recent study of the lower atmosphere based on photochemical modeling~\cite{Moses-2011} found that the molecular hydrogen in the atmosphere dissociates predominantly at pressures below about 1 microbar. Above this level, in the thermosphere, the gas is heated to high temperatures, since (i) the absorption of intense flows of extreme UV radiation from the parent star occurs there and (ii) due to the atomic struc- ture of the gas, no mechanisms for radiative cooling operate efficiently under these conditions. A recent model for the upper atmosphere of HD~209458b~\cite{Koskinen-2012} provides estimates of possible variations of the average (over the pressure) temperature of the upper atmosphere, resulting from calculations based on a hydrodynamical model. It was shown that, if the heating efficiency changes from 0.1 to 1.0, the average temperature in the upper atmosphere of HD~209458b varies from 6000 to 8000 K. Precisely this range of mean temperatures for the upper atmosphere of the ``hot Jupiter'' HD~209458b was used in calculations of the gas dynamics of the interaction of the neutral planetary atmosphere with the stellar wind.

The results of our 3D numerical simulations have confirmed the possible existence of all three types of atmospheres noted above. For the derived parameters of HD~209458b, all three types of envelopes can exist. Accordingly, since different types of envelopes have different observational manifestations, observational determinations of the type of envelope present could be used to impose additional constraints on the parameters of HD~209458b.

The applicability of the three types of envelopes considered to the various observed ``hot Jupiters'' and the criteria that can be used to discriminate between them are fundamental questions. The problem is that the analysis shown in Fig.~\ref{rhot}, which provided estimates of the critical density and temperature, is quite approximate, since it ignores a number of gas-dynamical effects. First and foremost, we did not take into account the possible deviation of the trajectory of the stream from a ballistic path. The approximation we have used is valid only if the density of the stream is much higher than the density of the ambient gas. If this is not the case, the HCP where the gas velocity and stream velocity are collinear could shift closer to or away from \Lp1, leading to a correction of the critical density~$\rho_0^*$. However, a comparison of our estimates and the calculation results shows that, due to the supersonic motion of the flow, gas-dynamical effects change the trajectory only slightly, requiring corrections of no more than a few percent. However, note that the interaction of the stream with the stellar wind can only lead to a shift of HCP down the $Y$ axis in Fig.~\ref{scheme}. This imposes a constraint on the onset of the eclipse caused by the stream.

Another important assumption is that various parameters of the stream, such as its temperature and cross-section, remain constant. The temperature can remain constant if the main source of heating is irradiation by the star. However, since there is a hot shock just ahead of the stream, the tempera- ture may increase as matter moves from \Lp1 to the HCP. The cross-section of the stream can also vary due to the compression of the stream by the stellar wind and the variation of the temperature. It is very difficult to take into account these and similar effects in analytical estimates, and simulations that correctly take into account the radiative transfer are no less difficult. Nevertheless, although the inclusion of radiative-transfer effects could result in significant changes of the basic parameters of the atmospheres, the criteria distinguishing the different types of atmospheres should, in general, remain the same.

Finally, our analysis has used a simple model for the stellar wind with zero tangential velocity, constant radial velocity, and the density decreasing proportional to the square of the distance from the star. These approximations may be valid for systems where the star rotates synchronously with the planetary orbit (or there is no substantial magnetic field to impart a rotational component to the stellar wind), and the stream stops fast enough so that we can neglect the gradient of the radial wind velocity. However, all three types of envelopes should also exist in this case, and the criteria distinguishing between them should not change significantly. Fundamental changes in the solution presented are possible only if the matter that has left the planetary atmosphere forms a torus-like envelope or disk. In this case, the problem becomes appreciably nonlinear (the gas of the atmosphere is added to the gas of the wind and changes the parameters from orbit to orbit), and the application of the above criteria becomes difficult. The influence of these effects requires further study. However, our assertion of the existence of three types of atmospheres also remains valid in this case.

\section*{Acknowledgments} 

This study was supported by the Basic-Research Programs of the Presidium of the Russian Academy of Sciences P-21 and P-22, the Russian Foundation for Basic Research (project nos. 11-02-00076, 12-02-00047, 13-02-00077, and 11-02-00479), the Program of Support for Leading Scientific Schools of the Russian Federation (grant NSh-3602.2012.2), the Ministry of Education and Science, and the Federal Targeted Program “Scientific and Science-Education Staff of Innovative Russia” for 2009– 2013.


\begin{thebibliography}{10}
\makeatletter\renewcommand\@biblabel[1]{#1.}\makeatother
\def\selectlanguageifdefined#1{
\expandafter\ifx\csname date#1\endcsname\relax
\else\language\csname l@#1\endcsname\fi}
\ifx\undefined\url\def\url#1{{\small #1}}\else\fi
\ifx\undefined\BibUrl\def\BibUrl#1{\url{#1}}\else\fi
\ifx\undefined\BibAnnote\long\def\BibAnnote#1{}\else\fi
\ifx\undefined\BibEmph\def\BibEmph#1{\emph{#1}}\else\fi

\bibitem{Murray-Clay-2009}
\selectlanguageifdefined{english}
R.~A.~{Murray-Clay}, E.~I.~{Chiang}, N.~{Murray}, \apj. {\bf 693}, 23 (2009).
\bibitem{Mayor-Queloz-1995}
\selectlanguageifdefined{english}
M.~{Mayor}, D.~{Queloz}, \nat. {\bf 378}, 355 (1995).
\bibitem{Charbonneau-2000}
\selectlanguageifdefined{english}
D.~{Charbonneau}, T.~M.~{Brown}, D.~W.~{Latham}, M.~{Mayor}, \apjl. {\bf 529},
  L45 (2000).
\bibitem{Vidal-Madjar-2003}
\selectlanguageifdefined{english}
A.~{Vidal-Madjar}, A.~{Lecavelier des Etangs}, J.-M.~{D{\'e}sert} et~al., \nat.
  {\bf 422}, 143 (2003).
\bibitem{Vidal-Madjar-2008}
\selectlanguageifdefined{english}
A.~{Vidal-Madjar}, A.~{Lecavelier des Etangs}, J.-M.~{D{\'e}sert} et~al.,
  \apjl. {\bf 676}, L57 (2008).
\bibitem{Ben-Jaffel-2007}
\selectlanguageifdefined{english}
L.~{Ben-Jaffel}, \apjl. {\bf 671}, L61 (2007).
\bibitem{Vidal-Madjar-2004}
\selectlanguageifdefined{english}
A.~{Vidal-Madjar}, J.-M.~{D{\'e}sert}, A.~{Lecavelier des Etangs} et~al.,
  \apjl. {\bf 604}, L69 (2004).
\bibitem{Ben-Jaffel-2009}
\selectlanguageifdefined{english}
L.~{Ben-Jaffel}, S.~{Sona Hosseini}, \apj. {\bf 709}, 1284 (2010).
\bibitem{Linsky-2010}
\selectlanguageifdefined{english}
J.~L.~{Linsky}, H.~{Yang}, K.~{France} et~al., \apj. {\bf 717}, 1291 (2010).
\bibitem{Yelle-2004}
\selectlanguageifdefined{english}
R.~V.~{Yelle}, \icarus. {\bf 170}, 167 (2004).
\bibitem{Munoz-2007}
\selectlanguageifdefined{english}
A.~{Garc{\'{\i}}a Mu{\~n}oz}, \planss. {\bf 55}, 1426 (2007).
\bibitem{Koskinen-2012}
\selectlanguageifdefined{english}
T.~T.~{Koskinen}, M.~J.~{Harris}, R.~V.~{Yelle}, P.~{Lavvas}, ArXiv e-prints
  (2012).
\bibitem{Lammer}
\selectlanguageifdefined{english}
H.~{Lammer}, K.~G.~{Kislyakova}, M.~{Holmstr{\"o}m} et~al., \apss. {\bf 335}, 9
  (2011).
\bibitem{Lecavelier-2010}
\selectlanguageifdefined{english}
A.~{Lecavelier Des Etangs}, D.~{Ehrenreich}, A.~{Vidal-Madjar} et~al., \aap.
  {\bf 514}, A72 (2010).
\bibitem{Fossati-et-al:2010}
\selectlanguageifdefined{english}
L.~{Fossati}, C.~A.~{Haswell}, C.~S.~{Froning} et~al., \apjl. {\bf 714}, L222
  (2010).
\bibitem{Fossati-et-al:2010b}
\selectlanguageifdefined{english}
L.~{Fossati}, S.~{Bagnulo}, A.~{Elmasli} et~al., \apj. {\bf 720}, 872 (2010).
\bibitem{Lai-et-al:2010}
\selectlanguageifdefined{english}
D.~{Lai}, C.~{Helling}, E.~P.~J.~{van den Heuvel}, \apj. {\bf 721}, 923 (2010).
\bibitem{Li-et-al:2010}
\selectlanguageifdefined{english}
S.-L.~{Li}, N.~{Miller}, D.~N.~C.~{Lin}, J.~J.~{Fortney}, \nat. {\bf 463}, 1054
  (2010).
\bibitem{Vidotto-2010}
\selectlanguageifdefined{english}
A.~A.~{Vidotto}, M.~{Jardine}, C.~{Helling}, \apjl. {\bf 722}, L168 (2010).
\bibitem{Vidotto-2011}
\selectlanguageifdefined{english}
A.~A.~{Vidotto}, M.~{Jardine}, C.~{Helling}, \mnras. {\bf 411}, L46 (2011).
\bibitem{Vidotto3}
\selectlanguageifdefined{english}
A.~A.~{Vidotto}, M.~{Jardine}, C.~{Helling}, \mnras. {\bf 414}, 1573 (2011).
\bibitem{Bisikalo-2013}
\selectlanguageifdefined{english}
D.~{Bisikalo}, P.~{Kaygorodov}, D.~{Ionov} et~al., \apj. {\bf 764}, 19 (2013).
\bibitem{Lecavelier-2012}
\selectlanguageifdefined{english}
A.~{Lecavelier des Etangs}, V.~{Bourrier}, P.~J.~{Wheatley} et~al., \aap. {\bf
  543}, L4 (2012).
\bibitem{Koskinen-2010}
\selectlanguageifdefined{english}
T.~T.~{Koskinen}, R.~V.~{Yelle}, P.~{Lavvas}, N.~K.~{Lewis}, \apj. {\bf 723},
  116 (2010).
\bibitem{Landau}
\selectlanguageifdefined{english}
L.~D.~Landau and E.~M.~Lifshits, \BibEmph{{Course of Theoretical Physics}}, Vol.~6: Fluid Mechanics (Nauka, Moscow, 1986; Pergamon, New York, 1987).
\bibitem{Baranov}
\selectlanguageifdefined{english}
V.~B.~Baranov and K.~V.~Krasnobaev, \BibEmph{{Hydrodynamical Theory of Cosmic Plasma}} (Nauka, Moscow, 1977) [in Russian].
\bibitem{Withbroe-et-al:1988}
\selectlanguageifdefined{english}
G.~L.~{Withbroe}, \apj. {\bf 325}, 442 (1988).
\bibitem{Lubow-Shu:75}
\selectlanguageifdefined{english}
S.~H.~{Lubow}, F.~H.~{Shu}, \apj. {\bf 198}, 383 (1975).
\bibitem{Boyarchuk-et-al:2002}
\selectlanguageifdefined{english}
A.~A.~{Boyarchuk}, D.~V.~{Bisikalo}, O.~A.~{Kuznetsov}, V.~M.~{Chechetkin},
  \BibEmph{{Mass transfer in close binary stars}} (2002).
\bibitem{Bisikalo-et-al:2003}
\selectlanguageifdefined{english}
D.~V.~Bisikalo, A.~A.~Boyarchuk, P.~V.~Kaigorodov, and
O.~A.~Kuznetsov, Astron. Rep. {\bf 47}, 809 (2003).
\bibitem{Moses-2011}
\selectlanguageifdefined{english}
J.~I.~{Moses}, C.~{Visscher}, J.~J.~{Fortney} et~al., \apj. {\bf 737}, 15
  (2011).
\end{thebibliography}

\end{document}